\documentstyle[epsfig,floats,aps,prl,twocolumn]{revtex}

\begin{document}
\author{W.-J. Huang and S.-C. Gou}
\address{Department of Physics,\\
National Changhua University of Education,\\
Changhua 50058, Taiwan}
\title{Field induced phase segregation and collective excitations of a trapped
spinor Bose-Einstein condensate }
\date{May 31, 1999}
\maketitle

\begin{abstract}
A hydrodynamic description is used to study the zero-temperature properties
of a trapped spinor Bose-Einstein condensate in the presence of a uniform
magnetic field. We show that, in the case of antiferromagnetic spin-spin
interaction, the polar and ferromagnetic configurations of the ground state
can coexist in the trap. These two phases are spatially segregated in such a
way that the polar state occupies the inner part while the ferromagnetic
state occupies the outer part of the atomic cloud. We also derive a set of
coupled hydrodynamic equations for the number density and spin density
excitations of the system. It is shown that these equations can be
analytically solved for the system in an isotropic harmonic trap and a
constant magnetic field. Remarkably, the related low lying excitation
spectra are completely determined by the solutions in the region occupied by
the polar state. We find that, within the Thomas-Fermi approximation, the
presence of a constant magnetic field does not change the excitation spectra
which still possess the similar form of that obtained by Stringari.
\end{abstract}

\pacs{03.75.Fi,05.30.Jp,32.80.Pj}

The recent realization of Bose-Einstein condensate (BEC) in trapped atomic
gases has inspired enormous interest in the theoretical and experimental
studies of inhomogeneous interacting Bose gases \cite{JILA,RICE,MIT}. More
recently, Stamper-Kurn {\it et} {\it al.} have successfully produced a BEC
of $^{23}$Na atoms in an optical dipole trap\cite{Stamper}. In their
experiment, a new kind of Bose-condensed system, namely, the spinor BEC
which is characterized by the three hyperfine spin states $\left|
F=1,m_{F}=\pm 1,0\right\rangle ,$ has been realized. A remarkable feature of
such a system is that, unlike the magnetically trapped BEC, the spin-flip
collisions between atoms allow population exchange among hyperfine states
without causing any trap loss. This has opened up the possibilities to
explore the Bose-condensed systems having internal degrees of freedom in
which the $U\left( 1\right) $ gauge symmetry as well as the rotational $%
SO\left( 3\right) $ symmetry in spin space are both involved.

In the spinor BEC, the spin-spin interaction between atoms can be
approximated by the spin exchange term ${\bf S}\cdot {\bf S}$ \cite{Ho}
where ${\bf S=}S_{x}{\bf e}_{x}+S_{y}{\bf e}_{y}+S_{z}{\bf e}_{z}$ is the
spin operator. In such circumstances, the sign of the spin-exchange coupling
strength may play a crucial role in determining the ground state
configurations. For example, in the cases of negative (ferromagnetic)
coupling strength, atoms of same spin orientation in the condensate tend to
attract each other against the repulsive binary scatterings which give rise
to the density-density interaction. The situation is equally complicated for
the cases of positive (antiferromagnetic) coupling strength in which atoms
of same spin orientation tend to repel each other. The competition between
the density-density and spin-spin interactions evidently lead to an
intriguing scenario of the spin dynamics of the spinor BEC that is revealed
by certain complex ground state structures \cite{Ho,Ohmi,Law,Huang}.

In the previous theoretical studies, the zero-temperature properties of the
spinor BEC have been considered exclusively either in a trap \cite{Ho,Law}
or in a uniform magnetic field \cite{Ohmi,Huang}. It is of great theoretical
interest to ask that how these physical properties are affected when the
confining trap and the external magnetic field are both present. In this
paper, we use the hydrodynamic treatment to probe into this issue with a
special emphasis on the case of antiferromagnetic spin-spin interaction. We
show that part of our results are related to the formation of spin domains
in the ground state of the spinor BEC which has been reported very recently
by Stenger {\it et} {\it al}. \cite{Stenger}.

We begin by considering the generalized Gross-Pitaeviski Hamiltonian 
\begin{eqnarray}
&&H=\int d^{3}r\left\{ \frac{1}{2m}\left| \nabla {\bf \Phi }\right|
^{2}+V\left( {\bf r}\right) \left| {\bf \Phi }\right| ^{2}-{\bf \Omega \cdot 
}\left( {\bf \Phi }^{\dagger }{\bf S\Phi }\right) \right.  \nonumber \\
&&\left. +\frac{1}{2}g_{n}\left| {\bf \Phi }\right| ^{4}+\frac{1}{2}%
g_{s}\left| {\bf \Phi }^{\dagger }{\bf S\Phi }\right| ^{2}\right\} ,\qquad
\left( \text{ }\hbar =1\right)  \label{Hamiltonian}
\end{eqnarray}
which describes the classical part of the spinor BEC \cite{Ho,Ohmi}. Here $%
V\left( {\bf r}\right) $ is the trapping potential and ${\bf \Omega }${\bf \ 
}is the Larmor frequency which is proportional to the magnetic field. For
simplicity, we shall assume that the magnetic field is applied along the $z$%
-axis, i.e., ${\bf \Omega }=\Omega _{L}{\bf e}_{z}.$ The coupling constants $%
g_{n}$ and $g_{s}$ characterizing the density-density and spin-spin
interaction, respectively, are given by $g_{n}=4\pi a_{n}/m$ and $g_{s}=4\pi
a_{s}/m$ , with $a_{n}$ and $a_{s}$ being the corresponding $s$-wave
scattering lengths. Unlike that for the scalar BEC, the order parameter $%
{\bf \Phi }$ for the present system is vector-like whose components are
represented by three classical fields, $\phi _{+},\phi _{0}$ and $\phi _{-},$
corresponding to the condensates in the hyperfine states $\left|
F=1,m_{F}=1,0,-1\right\rangle $, respectively. In the presence of a magnetic
field, there are two classes of symmetry transformations admitted by the
system, namely, the $U(1)$ phase symmetry and the $SO(2)$ rotational
symmetry about the direction of the field. It can be shown that the
underlying symmetry group of Eq.(\ref{Hamiltonian}) is $U\left( 1\right) 
\mathop{\textstyle \bigotimes }%
U\left( 1\right) $ such that the Hamiltonian is invariant under the
transformation: 
\begin{equation}
{\bf \Phi =}\left( \phi _{+},\phi _{0},\phi _{-}\right) ^{T}\rightarrow
\left( \phi _{+}e^{i\theta _{+}},\phi _{0}e^{i\theta _{0}},\phi
_{-}e^{i\theta _{-}}\right) ^{T}.  \label{U(1)*U(1)}
\end{equation}
Here the superscript $T$ stands for transpose and the parameterized phase
angles satisfy the relation $\theta _{+}+\theta _{-}=2\theta _{0}$.

We first examine the ground state configuration for the case of
ferromagnetic coupling ($g_{s}<0$), which is in essence identical with that
of a scalar BEC. In order to minimize the ground state energy, we apply the
inequality for spin-one systems, i.e., $\left| {\bf \Phi }^{\dagger }{\bf %
S\Phi }\right| \leq \left| {\bf \Phi }\right| ^{2}\equiv \rho \left( {\bf r}%
\right) $. Obviously, the contribution of all the spin-dependent parts in
Eq.(\ref{Hamiltonian}), 
\begin{equation}
-\Omega _{L}\left( {\bf \Phi }^{\dagger }S_{z}{\bf \Phi }\right) +\frac{1}{2}%
g_{s}\left| {\bf \Phi }^{\dagger }{\bf S\Phi }\right| ^{2},  \label{saber}
\end{equation}
is minimized by choosing ${\bf \Phi }^{\dagger }S_{z}{\bf \Phi }=\left| {\bf %
\Phi }\right| ^{2}.$ Now since ${\bf \Phi }^{\dagger }S_{z}{\bf \Phi =}%
\left| \phi _{+}\right| ^{2}-\left| \phi _{-}\right| ^{2}$ as we have
adopted the spin matrices in the $\left| m_{F}=1,0,-1\right\rangle $
representation, it follows immediately that $\phi _{0}=\phi _{-}=0$ and $%
\phi _{+}\neq 0$. Hence the ground state configuration based on the
hydrodynamic description is given by ${\bf \Phi }_{g}=\left( \sqrt{\rho _{0}(%
{\bf r)}}e^{i\theta _{0}({\bf r)}},0,0\right) ^{T}$, where $\rho _{0}$ and $%
\nabla \theta _{0}/m$ are identified as the local condensate density and
superfluid velocity, respectively. As we wish to deal with a system having
fixed total particle number $N=\int d^{3}r\rho _{0}\left( {\bf r}\right) ,$ $%
\rho _{0}$ and $\theta _{0}$ must be chosen to minimize the free energy 
\begin{eqnarray}
F\left[ {\bf \Phi }_{g}\right] &=&\int d^{3}r\left\{ \frac{1}{2m}\left(
\nabla \sqrt{\rho _{0}}\right) ^{2}+\frac{1}{2m}\rho _{0}\left( \nabla
\theta _{0}\right) ^{2}\right.  \nonumber \\
&&\left. +\left[ V\left( {\bf r}\right) -\Omega _{L}-\mu \right] \rho _{0}+%
\frac{1}{2}g_{1}\rho _{0}^{2}\right\} ,  \label{free-energy-A}
\end{eqnarray}
where $\mu $ is the chemical potential, and $g_{1}=g_{n}+g_{s}$. Obviously, $%
F\left[ {\bf \Phi }_{g}\right] $ can be minimized only when $\nabla \theta
_{0}=0,$ that is, $\theta _{0}$ must be position-independent. In the
Thomas-Fermi (TF) limit, the kinetic energy term $\left( \nabla \sqrt{\rho
_{0}}\right) ^{2}/2m$ is neglected and the minimization of the free energy
can be achieved by letting 
\begin{equation}
\rho _{0}=\left\{ 
\begin{array}{ll}
g_{1}^{-1}\left[ \mu +\Omega _{L}-V\left( {\bf r}\right) \right] , & \mu
+\Omega _{L}\geq V\left( {\bf r}\right) \\ 
0, & \text{elsewhere}
\end{array}
\right. .  \label{density-ferro}
\end{equation}
We next consider the collective motion of the order parameter. In doing so,
we assume that the collective modes deviate slightly from the ground state
by $\delta {\bf \Phi }\left( {\bf r},t\right) $, i.e., ${\bf \Phi }\left( 
{\bf r},t\right) ={\bf \Phi }_{g}\left( {\bf r}\right) +\delta {\bf \Phi }%
\left( {\bf r},t\right) $. Substituting ${\bf \Phi }\left( {\bf r},t\right) $
into the motion equation of the system 
\begin{eqnarray}
&&i\frac{\partial }{\partial t}{\bf \Phi }=\left[ -\frac{\nabla ^{2}}{2m}%
+V\left( {\bf r}\right) -\mu -\Omega _{L}S_{z}\right] {\bf \Phi }  \nonumber
\\
&&+g_{n}\left| {\bf \Phi }\right| ^{2}{\bf \Phi }+g_{s}\left( {\bf \Phi }%
^{\dagger }{\bf S\Phi }\right) \cdot {\bf S\Phi ,}  \label{TDGP-eq}
\end{eqnarray}
it is easy to check that the small fluctuation parts $\delta \phi _{0}${\bf %
\ }and{\bf \ }$\delta \phi _{-}$ lead to two independent modes whose spectra
are determined by the one-particle Schr\"{o}dinger equations with effective
potentials $V\left( {\bf r}\right) +g_{1}\rho _{0}-\mu $ and $V\left( {\bf r}%
\right) +\left( g_{n}-g_{s}\right) \rho _{0}+\Omega _{L}-\mu $,
respectively. In the homogeneous case, each mode is known to have a
free-particle-like spectrum with a gap\cite{Ohmi,Huang}. The
hydrodynamic(-like) mode corresponding to the Bose-condensed component $\phi
_{+}$ can be obtained by considering order parameter of the form 
\begin{equation}
{\bf \Phi }\left( {\bf r},t\right) =\left( \sqrt{\rho \left( {\bf r}%
,t\right) }e^{i\theta \left( {\bf r},t\right) },0,0\right) ^{T},
\label{coll-fero}
\end{equation}
where $\rho \left( {\bf r},t\right) =\rho _{0}\left( {\bf r}\right) +\delta
\rho \left( {\bf r},t\right) .$ Substituting Eq.(\ref{coll-fero}) into Eq.(%
\ref{TDGP-eq}) and following the analytical approach given by Stringari \cite
{Stringari}, it is straightforward to show that the density fluctuation $%
\delta \rho $ satisfies the equation of motion 
\begin{equation}
\frac{\partial ^{2}}{\partial t^{2}}\delta \rho =\frac{g_{1}}{m}\nabla \cdot
\left[ \rho _{0}\nabla \delta \rho \right] .  \label{Stringari}
\end{equation}
Accordingly, the low lying collective excitation spectrum is identical with
that of the universal excitation spectrum obtained by Stringari\cite
{Stringari}.

Of particular interest is the antiferromagnetic case $(g_{s}>0).$ Clearly,
the two terms given by Eq.(\ref{saber}) are minimized only when ${\bf \Phi }%
^{\dagger }S_{x}{\bf \Phi }={\bf \Phi }^{\dagger }S_{y}{\bf \Phi }=0$ or,
explicitly, $\phi _{+}^{*}\phi _{0}+\phi _{0}^{*}\phi _{-}=\phi _{0}^{*}\phi
_{+}+\phi _{-}^{*}\phi _{0}=0$. Furthermore, the minimum is achieved if we
impose that ${\bf \Phi }^{\dagger }S_{z}{\bf \Phi }=\Omega _{L}/g_{s}>0$.
When $\phi _{0}\neq 0,$ by using Eq.(\ref{U(1)*U(1)}), we may assume without
loss of generality that $\phi _{0}$ is real. This assumption implies that $%
\phi _{+}^{*}=-\phi _{-}$ and, as a consequence, ${\bf \Phi }^{\dagger }S_{z}%
{\bf \Phi }=0$ which apparently violates the imposed condition ${\bf \Phi }%
^{\dagger }S_{z}{\bf \Phi }=\Omega _{L}/g_{s}.$ We thus conclude that $\phi
_{0}$ must be identically equal to zero and are led to the so-called polar
state for the condensate\cite{Ho}: ${\bf \Phi }_{g}^{\left( P\right) }\left( 
{\bf r}\right) =\left( \left| \phi _{+}^{\left( P\right) }\right| e^{i\theta
_{+0}},0,\left| \phi _{-}^{\left( P\right) }\right| e^{i\theta _{-0}}\right)
^{T}$, where $\left| \phi _{\pm }^{\left( P\right) }\right| ^{2}=\rho
_{0}\left( 1\pm \Omega _{L}/\rho _{0}g_{s}\right) /2$. Note that in this
region the condition $\rho _{0}\left( {\bf r}\right) \geq \Omega _{L}/g_{s}$
is required in order to keep $\left| \phi _{-}^{\left( P\right) }\right|
^{2}\geq 0$. On the other hand, if $\rho _{0}\left( {\bf r}\right) <\Omega
_{L}/g_{s}$, the condition ${\bf \Phi }^{\dagger }S_{z}{\bf \Phi }=\Omega
_{L}/g_{s}>0$ can never be fulfilled and we must choose ${\bf \Phi }$ $\ $in
such a way that ${\bf \Phi }^{\dagger }S_{z}{\bf \Phi }$ is as close to $%
\Omega _{L}/g_{s}$ as possible. In view of the inequality, $\left| {\bf \Phi 
}^{\dagger }{\bf S\Phi }\right| \leq \left| {\bf \Phi }\right| ^{2}$, we
conclude that the ground state configuration is described by the
ferromagnetic state\cite{Ho}: ${\bf \Phi }_{g}^{\left( F\right) }\left( {\bf %
r}\right) =\left( \rho _{0}e^{i\theta _{+0}},0,0\right) ^{T}$. We thus see
that when $g_{s}>0$, the two different ground state configurations ${\bf %
\Phi }_{g}^{\left( P\right) }$ and ${\bf \Phi }_{g}^{\left( F\right) }$
coexist and the phase boundary ${\bf r}_{b}$ is determined by $\rho
_{0}\left( {\bf r}_{b}\right) =\Omega _{L}/g_{s}.$ The free energy of such a
structure of coexistence can be described by 
\begin{eqnarray}
F\left[ {\bf \Phi }_{g}\right]  &=&\int\limits_{\text{polar }}d^{3}r\left[ 
\frac{1}{2m}\left( \nabla \sqrt{\rho _{0}}\right) ^{2}\frac{1}{1-\left(
\Omega _{L}/\rho _{0}g_{s}\right) ^{2}}-\frac{\Omega _{L}^{2}}{2g_{s}}%
\right]   \nonumber \\
&&+\int\limits_{\text{ferro}}d^{3}r\left[ \frac{1}{2m}\left( \nabla \sqrt{%
\rho _{0}}\right) ^{2}-\Omega _{L}\rho _{0}+\frac{1}{2}g_{s}\rho
_{0}^{2}\right]   \nonumber \\
&&+\int d^{3}r\left[ \left( V\left( {\bf r}\right) -\mu \right) \rho _{0}+%
\frac{1}{2}g_{n}\rho _{0}^{2}\right] ,  \label{total-free-energy}
\end{eqnarray}
where we have set $\nabla \theta _{+0}\left( {\bf r}\right) =\nabla \theta
_{-0}\left( {\bf r}\right) =0$ which serves as the necessary condition for
minimizing the free energy. We subsequently consider the minimization
condition, $\delta F/\delta \rho _{0}=0,$ for both regions$.$ In the TF
limit, the kinetic energy term is ignored and hence we are able to write
down the density profile in both regions: 
\begin{equation}
\rho _{0}\left( {\bf r}\right) =\left\{ 
\begin{array}{ll}
g_{n}^{-1}\left[ \mu -V\left( {\bf r}\right) \right] , & \text{polar ;} \\ 
g_{1}^{-1}\left[ \mu +\Omega _{L}-V\left( {\bf r}\right) \right] , & \text{%
ferromagnetic.}
\end{array}
\qquad \right.   \label{density-profile}
\end{equation}
For simplicity, we assume an isotropic harmonic trap $V_{har}\left( r\right)
=m\omega _{0}^{2}r^{2}/2.$ Thus, the two important length scales of the
density profile Eq.(\ref{density-profile}), namely, the phase boundary and
the radius of the atomic cloud are given by $r_{b}=\omega _{0}^{-1}\left[
2\left( \mu -\Omega _{L}g_{n}/g_{s}\right) /m\right] ^{1/2}$ and $R=\omega
_{0}^{-1}\left[ 2\left( \mu +\Omega _{L}\right) /m\right] ^{1/2}$,
respectively. In this case, one sees that the $\phi _{-}$ component is
restricted in the region of $0\leq r\leq r_{b}$ while the $\phi _{+}$
component exists ubiquitously in the atomic cloud. We note that a similar
structure has been reported in Refs.\cite{Stenger,Isoshima}, in the case
that the quadratic Zeeman energy is almost canceled. The chemical potential
can be obtained by fixing the total particle number with the density profile
led by $V_{har}\left( r\right) $: 
\begin{eqnarray}
N &=&\frac{8\pi }{15g_{1}}\left( \frac{2}{\omega _{0}^{2}m}\right)
^{3/2}\left[ \frac{g_{s}}{g_{n}}\left( \mu -\frac{g_{n}}{g_{s}}\Omega
_{L}\right) ^{5/2}\right.   \nonumber \\
&&+\left. \left( \mu +\Omega _{L}\right) ^{5/2}\right] .
\end{eqnarray}
This equation provides a complicated mathematical relation between the
chemical potential $\mu $ and the total particle number $N$. However, when
the magnetic field is absent, we again obtain the well-known 5/2-power-law
for the trapped scalar BEC: $N=a_{HO}\left( 2\mu /\omega _{0}\right)
^{5/2}/15a_{n}$, where $a_{HO}=\left( m\omega _{0}\right) ^{-1/2}$ is the
characteristic length of the harmonic trap.

In order to obtain the low lying excitation spectrum of the configuration
described in Eq.(\ref{density-profile}), we let 
\begin{equation}
{\bf \Phi }\left( {\bf r,}t\right) =\left\{ 
\begin{array}{ll}
\left( \sqrt{\rho }e^{i\theta _{+}},\delta \phi _{0},\sqrt{\rho _{-}}%
e^{i\theta _{-}}\right) ^{T}, & \text{polar} \\ 
\left( \sqrt{\rho }e^{i\theta _{+}},\delta \phi _{0},\delta \phi _{-}\right)
^{T},\quad  & \text{ferromagnetic}
\end{array}
\right. 
\end{equation}
Here $\rho _{\pm }\left( {\bf r,}t\right) =\rho _{0\pm }\left( {\bf r}%
\right) +\delta \rho _{\pm }\left( {\bf r,}t\right) $, where $\rho _{0\pm
}\left( {\bf r}\right) $ denote the equilibrium particle densities for the $%
\phi _{\pm }$ fields in the configuration of Eq.(\ref{density-profile}) and $%
\delta \rho _{\pm }\left( {\bf r,}t\right) $ are the corresponding density
fluctuations. In the ferromagnetic region the respective linearized
equations for the fluctuations $\delta \rho _{+},\delta \phi _{0},$ and $%
\delta \phi _{-}$ are again decoupled and are, in fact, exactly the same as
those for the case of ferromagnetic coupling. In the polar region, the
fluctuation $\delta \phi _{0}$ also decouples from $\delta \rho _{\pm }$. In
fact, we have derived the corresponding linearized equation and found that
it is not analytically solvable in the presence of a parabolic trapping
potential. Here we shall not discuss that equation which is beyond the scope
of this paper. Nevertheless, it would be instructive to point out that in
the homogenous case, the motion equation of $\delta \phi _{0}$ is solvable
and predicts the existence of a massive mode with a gap equal to $\Omega _{L}
$\cite{Ohmi,Huang}. Again, we follow the steps in deriving Eq.(\ref
{Stringari}) and obtain the following linearized coupled equations: 
\begin{eqnarray}
\frac{\partial ^{2}\delta \rho _{+}}{\partial t^{2}} &=&\nabla \cdot \left\{ 
\frac{\rho _{0+}}{m}\left[ g_{1}\nabla \delta \rho _{+}+\left(
g_{n}-g_{s}\right) \nabla \delta \rho _{-}\right] \right\} ,  \nonumber \\
\frac{\partial ^{2}\delta \rho _{-}}{\partial t^{2}} &=&\nabla \cdot \left\{ 
\frac{\rho _{0-}}{m}\left[ \left( g_{n}-g_{s}\right) \nabla \delta \rho
_{+}+g_{1}\nabla \delta \rho _{-}\right] \right\} .  \label{ssd}
\end{eqnarray}
Let $\delta \rho =\delta \rho _{+}+\delta \rho _{-}$ and $\delta \rho
_{s}=\delta \rho _{+}-\delta \rho _{-}$ which correspond to the fluctuations
of number density and spin density respectively. If we further assume that
the time dependence of these two quantities are described by $\delta \rho
\left( {\bf r},t\right) =\delta \rho \left( {\bf r}\right) \exp \left(
-i\omega t\right) $ and $\delta \rho _{s}\left( {\bf r},t\right) =\delta
\rho _{s}\left( {\bf r}\right) \exp \left( -i\omega t\right) $, we then
obtain the desired hydrodynamic equations for the system: 
\begin{eqnarray}
\omega ^{2}\delta \rho  &=&-\nabla \cdot \left\{ g_{n}\frac{\rho _{0}}{m}%
\nabla \delta \rho +\frac{\Omega _{L}}{m}\nabla \delta \rho _{s}\right\}  
\nonumber \\
\omega ^{2}\delta \rho _{s} &=&-\nabla \cdot \left\{ \frac{g_{n}}{g_{s}}%
\frac{\Omega _{L}}{m}\nabla \delta \rho +g_{s}\frac{\rho _{0}}{m}\nabla
\delta \rho _{s}\right\} .  \label{hydro-equation}
\end{eqnarray}
We note that when the magnetic field is absent, Eq.(\ref{hydro-equation})
reduces to the same result obtained by Ho \cite{Ho}. On the other hand, if
the trap is removed, the density $\rho _{0}$ becomes a constant and it is
thus easy to see that the excitation frequencies agree with the results
given by Ohmi and Machida in the long-wavelength limit\cite{Ohmi}.

In order to solve Eq.(\ref{hydro-equation}) for an isotropic harmonic
potential, it is useful to assume solutions of the form: 
\begin{equation}
\delta \rho \left( {\bf r}\right) =f\left( r\right) Y_{lm}\left( \theta
,\phi \right) ,\delta \rho _{s}\left( {\bf r}\right) =f_{s}\left( r\right)
Y_{lm}\left( \theta ,\phi \right) ,
\end{equation}
where $Y_{lm}\left( \theta ,\phi \right) $ is the spherical harmonic
function. In what follows, we merely substitute the density profile for $%
V_{har}\left( r\right) $ in the polar region into Eq.(\ref{hydro-equation})
as it will be illustrated later that this density profile alone is
sufficient to solve the excitation spectrum of the system. Introducing the
dimensionless variable $x\equiv r/R,$ the coupled equation for $f\left(
x\right) $ and $f_{s}\left( x\right) $ in the polar region $\left( 0\leq
x\leq r_{b}/R\right) $ is expressed by the following matrix equation 
\begin{eqnarray}
{\bf A}\left( \frac{d^{2}}{dx^{2}}+\frac{2}{x}\frac{d}{dx}-\frac{l\left(
l+1\right) }{x^{2}}\right) {\bf f} &&  \label{matrix-equation} \\
-\left[ x^{2}\frac{d^{2}}{dx^{2}}+4x\frac{d}{dx}-l\left( l+1\right) \right] 
{\bf f}+{\bf Bf} &=&0,  \nonumber
\end{eqnarray}
where the matrices are given by 
\[
{\bf A}=\left( 
\begin{array}{cc}
1 & x_{L}^{2} \\ 
x_{L}^{2}/\beta ^{2} & 1
\end{array}
\right) ,{\bf B}=\left( 
\begin{array}{cc}
\epsilon & 0 \\ 
0 & \epsilon /\beta
\end{array}
\right) ,{\bf f}=\left( 
\begin{array}{c}
f\left( x\right) \\ 
f_{s}\left( x\right)
\end{array}
\right) , 
\]
with the parameters defined by $\epsilon =2\omega ^{2}/\omega _{0}^{2},$ $%
x_{L}^{2}=2\omega _{L}/m\omega _{0}^{2}R^{2}$ and $\beta =g_{s}/g_{n}.$
Equation (\ref{matrix-equation}) can be solved by using the series
expansions for the radial functions $f\left( x\right) $ and $f_{s}\left(
x\right) $, i.e., ${\bf f}=\sum\limits_{k=0}^{\infty }{\bf u}_{k}x^{k+l}$
where ${\bf u}_{k}=\left( a_{k},b_{k}\right) ^{T}$. It is straightforward to
show that for $k\geq -1$, we have, with the convention ${\bf u}_{-1}=0$, the
recursion relation 
\begin{equation}
{\bf Au}_{k+2}=-\frac{{\bf B}-k^{2}-2kl-3k-2l}{\left( k+2\right) \left(
k+2l+3\right) }{\bf u}_{k}.  \label{recursion}
\end{equation}
An immediate consequence of Eq.(\ref{recursion}) is that ${\bf u}_{k}$
vanishes for all odd $k$. To solve Eq.(\ref{recursion}) for even $k$ we
first work out the eigenvalues $\lambda _{\pm }$ of the matrix ${\bf A}$
which are given by $\lambda _{\pm }=1\pm x_{L}^{2}/\beta $. Let ${\bf u}%
_{\pm }$ be the corresponding eigenvectors of $\lambda _{\pm }$ and thus for
all $k$ we have ${\bf u}_{k}=c_{k}^{\left( +\right) }{\bf u}%
_{+}+c_{k}^{\left( -\right) }{\bf u}_{-}$. Accordingly, Eq.(\ref{recursion})
becomes 
\begin{eqnarray}
&&\lambda _{+}c_{k+2}^{\left( +\right) }{\bf u}_{+}+\lambda
_{-}c_{k+2}^{\left( -\right) }{\bf u}_{-}  \nonumber \\
&=&-\frac{{\bf B}-k^{2}-2kl-3k-2l}{\left( k+2\right) \left( k+2l+3\right) }%
\left( c_{k}^{\left( +\right) }{\bf u}_{+}+c_{k}^{\left( -\right) }{\bf u}%
_{-}\right) .
\end{eqnarray}
When $\beta \neq 1\left( g_{n}\neq g_{s}\right) $, it is obvious that ${\bf u%
}_{\pm }$ can not be eigenvectors of ${\bf B}$. A close examination on the
behavior of the series in the large order expansion shows that the series
must terminate at some $k$, otherwise it will diverge at the phase boundary $%
r=r_{b}$ . We therefore conclude that for some $k=2n\geq 0,$ the right hand
side of Eq.(\ref{recursion}) vanishes and thus the solutions to the
eigenvalue problems are 
\begin{equation}
{\bf u}_{2n}=\left\{ 
\begin{array}{ll}
\left( a_{2n},0\right) ^{T}, & \epsilon =4n^{2}+4nl+6n+2l; \\ 
\left( 0,b_{2n}\right) ^{T}, & \epsilon =\beta \left( 4n^{2}+4nl+6n+2l\right)
\end{array}
\right.  \label{eigen-system}
\end{equation}
In either case, the sequence of the vectors ${\bf u}_{k}$ can be determined
by the use of Eq.(\ref{recursion}) and thus the corresponding eigenfunctions 
${\bf f}$ can be obtained. For $\beta =1$, the situation is simpler: Eq.(\ref
{ssd}) becomes two independent equations for $\delta \rho _{\pm }$. A
similar analysis shows that Eq.(\ref{eigen-system}) is again valid.

The first mode in Eq.(\ref{eigen-system}) has the spectrum $\omega
_{nl}^{\left( 1\right) }=\omega _{0}\left( 2n^{2}+2nl+3n+l\right) ^{1/2}$
which possesses exactly the same form of the universal spectrum obtained by
Stringari \cite{Stringari}. What is noticeable is the spectrum of the second
mode $\omega _{nl}^{\left( 2\right) }=\omega _{0}\left[ \left(
2n^{2}+2nl+3n+l\right) g_{s}/g_{n}\right] ^{1/2}$ which now depends
explicitly on the two-body interaction although it has the same $n,l$
dependence as that of $\omega _{nl}^{\left( 1\right) }.$ When the magnetic
field is absent the two modes are merely owing to the number density and the
spin density fluctuations, respectively. It should be noted that both two
sets of dispersion relations given by Eq.(\ref{eigen-system}) do not depend
on the magnetic field at all. In other words, the presence of a constant
magnetic field does not change the frequencies of the hydrodynamic modes but
indeed mixes up the number density oscillation $\delta \rho $ and the spin
density oscillation $\delta \rho _{s}$. It is also interesting to point out
that when $g_{n}=g_{s},$ we have two independent hydrodynamic equation of
the Stringari-type which have the same spectra.

In conclusion, we have investigated the ground state structure of a trapped
spinor BEC in the presence of a uniform magnetic field. It is found that the
antiferromagnetic spin-spin interaction leads to the coexistence of the
polar and ferromagnetic phases in the ground state. For the isotropic
harmonic trap the density profile is obtained in the TF limit. Moreover, the
linearized equations for the collective modes are derived. We find that the
oscillation frequencies of the hydrodynamic modes are completely determined
by the linearized equations in the polar region. Within the TF
approximation, the spectra of the hydrodynamic modes are shown to be
independent of the strength of the magnetic field and possess a form of
Stringari-type.

Our studies provide a theoretical treatment for the zero-temperature
properties of the trapped spinor BEC. These analyses are, however, limited
by the TF approximation and the use of an isotropic trap. For all practical
purposes, it is necessary to extend the present calculations to include the
effects of the kinetic energy pressure as well as the anisotropic trap.

This work is supported by the National Science Council, Taiwan under Grant
NSC-88-2112--M-018-004.

\end{document}